# Self-formed 2D/3D Heterostructure on the Edge of 2D Ruddlesden-Popper Hybrid Perovskites Responsible for Intriguing Optoelectronic Properties and Higher Cell Efficiency

Zhaojun Qin[1,2], Shenyu Dai[2,3], Chalapathi Charan Gajjala[2], Chong Wang[4], Viktor G. Hadjiev[5], Guang Yang[6], Jiabing Li[2], Xin Zhong[5,7], Zhongjia Tang[5,7], Yan Yao[2], Arnold M. Guloy[5,7], Rohith Reddy[2], David Mayerich[2], Liangzi Deng[5,8], Qingkai Yu[9], Guoying Feng[3], Zhiming Wang[1,]*, Jiming Bao[2,6,]*

[1]Institute of Fundamental and Frontier Sciences, University of Electronic Science and Technology of China, Chengdu, Sichuan 610054, China

[2]Department of Electrical and Computer Engineering, University of Houston, Houston, Texas 77204, USA

[3]College of Electronics & Information Engineering, Sichuan University, Chengdu, Sichuan 610064, China

[4]School of Materials and Energy, Yunnan University, Kunming, Yunnan 650091, China

[5]Texas Center for Superconductivity, University of Houston, Houston, TX 77204, USA

[6]Materials Science & Engineering, University of Houston, Houston, Texas 77204, USA

[7]Department of Chemistry, University of Houston, Houston, Texas 77204, USA

[8]Department of Physics, University of Houston, Houston, Texas 77204, USA

[9]Ingram School of Engineering, Texas State University, San Marcos, Texas 78666, USA

*To whom correspondence should be addressed: Jiming Bao (jbao@uh.edu),

Zhiming Wang (zhmwang@uestc.edu.cn)

# Abstract

The observation of low energy edge photoluminescence and its beneficial effect on the solar cell efficiency of Ruddlesden-Popper perovskites has unleashed an intensive research effort to reveal its origin. This effort, however, has been met with more challenges as the underlying material structure has still not been identified; new modellings and observations also do not seem to converge. Using 2D $(BA)_2(MA)_2Pb_3Br_{10}$ as an example, we show that 3D $MAPbBr_3$ is formed due to the loss of BA on the edge. This self-formed $MAPbBr_3$ can explain the reported edge emission under various conditions, while the reported intriguing optoelectronic properties such as fast exciton trapping from the interior 2D perovskite, rapid exciton dissociation and long carrier lifetime can be understood via the self-formed 2D/3D lateral perovskite heterostructure. The 3D perovskite is identified by submicron infrared spectroscopy, the emergence of XRD signature from freezer-milled nanometer-sized 2D perovskite and its photoluminescence response to external hydrostatic pressure. The revelation of this edge emission mystery and the identification of a self-formed 2D/3D heterostructure provide a new approach to engineering 2D perovskites for high-performance optoelectronic devices.

Two-dimensional (2D) Ruddlesden-Popper (RP) hybrid perovskites are stacks of atomically thin layers of 3D perovskites separated by organic long-chain cations. Because of their organic spacers leading to a higher material stability than that of their 3D counterparts[1, 2], RP perovskites have attracted a lot of attention and have been used to fabricate high-performance solar cells, light-emitting diodes and photodetectors[1, 2, 3, 4, 5, 6, 7, 8, 9, 10, 11, 12, 13]. RP perovskites are also an ideal platform for people to explore fundamental physics and do bandgap engineering: by tuning the thickness of layers, the bandgap and exciton binding energies can be tuned as in semiconductor quantum wells and superlattices[2, 3, 4, 13]. Recent surprising observations of intriguing edge photoluminescence and its benefits to solar cells have attracted more attention over past several years[14]. However, the microscopic structure of potential edge state has not been identified and the origin of the edge emission remains controversial despite many efforts from both theoretical and experimental groups[15, 16, 17, 18, 19, 20].

The mystery began in 2017 when Blancon *et al.* reported a potential edge state which exhibited a lower energy photoluminescence (PL) than that of exfoliated RP perovskites $(BA)_2(MA)_{n-1}Pb_nI_{3n+1}$ on its edge when n≥3[14]. This edge state is further shown to trap photo-excited excitons from the interior of the perovskite and dissociate them into long-lived free carriers. By orienting edges of RP perovskites along direction of photocurrent, a higher solar cell efficiency was achieved[14]. This finding was very surprising, because edges, like grain boundaries, break crystal symmetry, and they are typically the sources of defect states and are detrimental to the device performances. However, the origin of the edge photoluminescence and its underlying microstructure were not understood and identified[14]. After that, five related papers have been published to either try to solve this mystery or to report new interesting physics[15, 16, 17, 18, 19], but their results do not converge, making the puzzle more difficult to solve. Two of them are theoretical papers, and they proposed very different models. Blancon and his collaborators first proposed that the edge emission and edge state are induced by strain relaxation only when n≥3[15]. Later Zhang *et al.*, however, proposed that it is the difference in chemistry between Pb and iodine on the edge that leads to exciton dissociation and localization[18]. In both cases, the edge states are an intrinsic property of 2D perovskites. Experiments, however, point to the opposite direction and suggesting that the edge emissions are extrinsic and can be controlled by external environments or chemistry. In 2019, Shi *et al*. reported that the edge emission was induced by moisture, they further observed the edge emission when n=2 in different perovskites such as $BA_2FAPb_2Br_7$ and $BA_2FAPb_2I_7$, in conflict with earlier observation and theory[16]. Late in the same year, Zhao et al. reported the control of the edge emission by BAI and MAI[17]. Very recently, Wang *et al*. discovered that the edge has a much higher electrical conductivity than the center region[19].

It is not surprising that the origin of the edge emission remains mysterious despite these intensive efforts. The main reason is the lack of identification of the underlying structure[20]. Although TEM was employed and found some differences between the edge and the center, no definite structure was identified[16]. Certainly, the large discrepancies among different theories and observations described above have made it more difficult to resolve this mystery[15, 16, 17, 18, 19]. In this work, we report the resolution of this mystery and identification of underlying structure: it is the 3D perovskite and the 2D/3D perovskite heterostructure that are responsible for the lower energy edge emission and many intriguing and beneficial optoelectronic properties. 3D perovskite is naturally formed after the loss or replacement of BA and subsequent connection of lead halide octahedrons in the vertical direction. Its identification is made possible by a combination of conventional techniques and novel approaches which include XRD of freezer milled 2D perovskite, infrared spectroscopy at sub-micrometer scale, and characteristic response of photoluminescence to hydrostatic pressure in a diamond anvil cell (DAC)[21, 22].

**Results and Discussion**

As the first piece of evidence, we noticed that the reported edge PL peaks in the exfoliated 2D perovskites are always similar to those of the corresponding 3D perovskites and is independent of the layer thickness n. For example, the PL peak in the edge of $(BA)_2(MA)_{n-1}Pb_nI_{3n+1}$ (n=3-5) is centered at ~740 nm (1.68 eV)[14, 16, 17], close to that of the bulk PL in 3D $MAPbI_3$ (~770 nm)[23, 24, 25, 26]. Exfoliated 2D $(BA)_2(MA)_{n-1}Pb_nBr_{3n+1}$ (n=2, 3) exhibited edge emission at ~520 nm[16], while PL of the corresponding 3D $MAPbBr_3$ is centered at ~550 nm[27, 28]. The edge PL of $(BA)_2FAPb_2I_7$ and $(BA)_2FAPb_2Br_7$ are centered at 750 nm and 525 nm respectively, very close to 800 nm and 540 nm of their 3D counterparts[29, 30, 31].

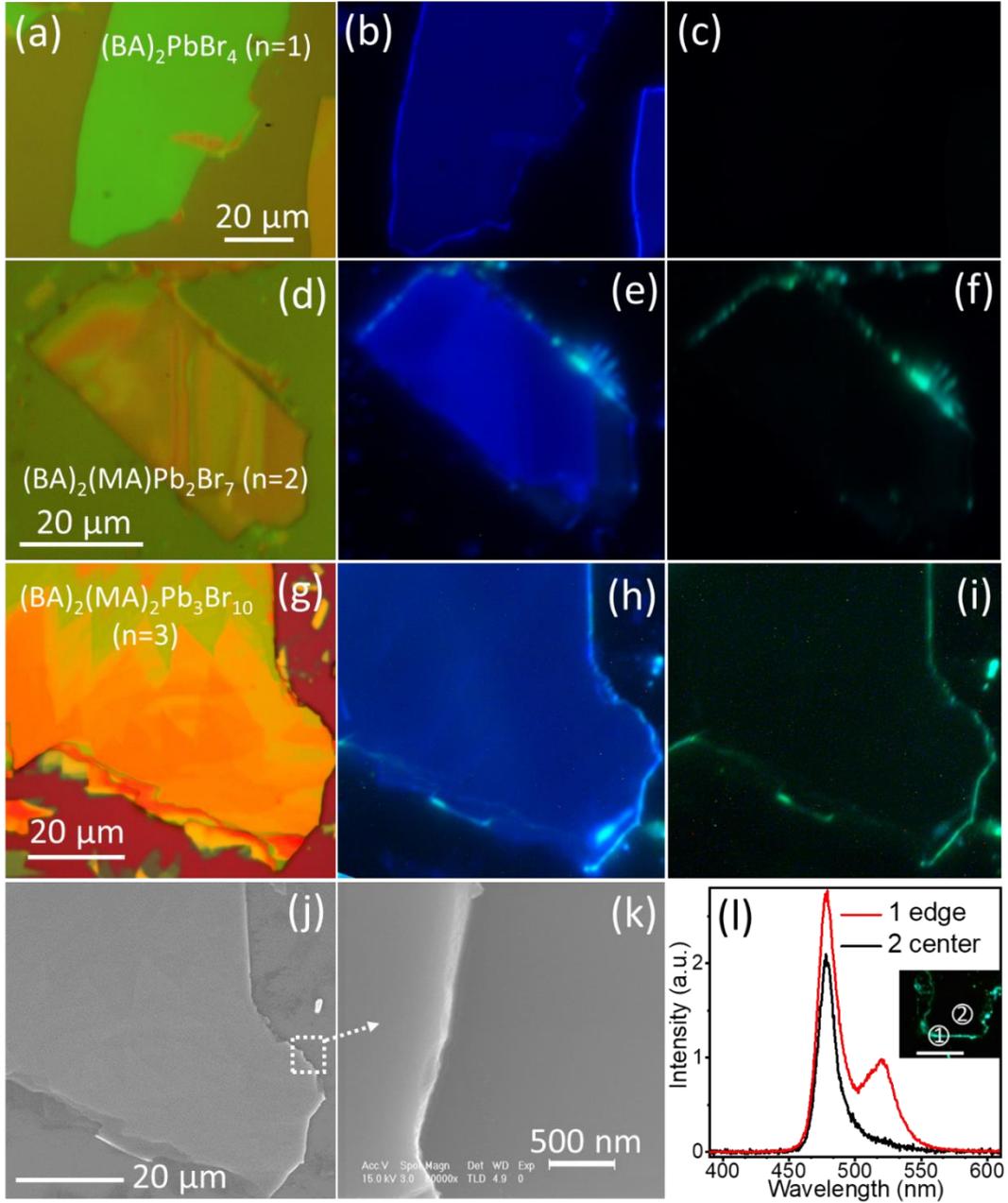

Figure 1. (a-i) Optical images (a, d, g), 365-nm UV-excited PL images with 420 nm (b, e, h) and 500 nm (c, f, i) long-pass filters of exfoliated $(BA)_2(MA)_{n-1}Pb_nBr_{3n+1}$ for n=1 (a-c), n=2 (d-f), and n=3 (g-i). (j-k) Low (j) and high (k) magnification SEM images of the n=3 perovskite in (g-i). (l) Representative PL spectra of n=3 perovskite from the edge and center regions as indicated in the PL image in the inset.

To provide our own evidence, we synthesized representative RP perovskites $(BA)_2(MA)_{n-1}Pb_nBr_{3n+1}$ (n=1, 2, 3) from a stoichiometric reaction between PbO, MABr and BABr[4, 16]. The PL images and peak positions in Figure 1 confirm the edge emission for n=2 and 3 only, in agreement with previous report[16]. Scanning electron microscopy (SEM) images in Figures 1j-k also confirm that emissive edges $(BA)_2(MA)_2Pb_3Br_{10}$ are smooth, as opposed to rough edge in $CsPb_2Br_5$ platelets, suggesting that the edge emission does not come from overgrown or contaminated nanostructures during the synthesis[21]. To determine whether the edge emission is an intrinsic or extrinsic property of the RP perovskites, we created fresh edges from a large piece of $(BA)_2(MA)_2Pb_3Br_{10}$ using the tip of a probe, and monitored its edge emission. Figure S1 shows that the PL emission from the fresh edges became stronger over time in air, indicating that the edge emission is not an intrinsic property as proposed by two theory papers[15, 18]. In the following experiments, we will use $(BA)_2(MA)_2Pb_3Br_{10}$ to prove the existence of 3D $MAPbBr_3$ on the edge.

X-ray diffraction (XRD) is the most reliable technique to confirm and identify a material or structure, however, a direct X-ray survey in figure S2 shows no sign of 3D $MAPbBr_3$ in the as-grown $(BA)_2(MA)_2Pb_3Br_{10}$. This is because the sensitivity of XRD is too low to detect trace amount of 3D structure on the edge[20]. This problem can be overcome by making smaller size 2D perovskites so that the surface (edge) to volume ratio can be significantly increased. Here we chose a freezer mill at liquid nitrogen temperature to physically break large bulk crystals to smaller disks without causing any chemical or compositional changes.

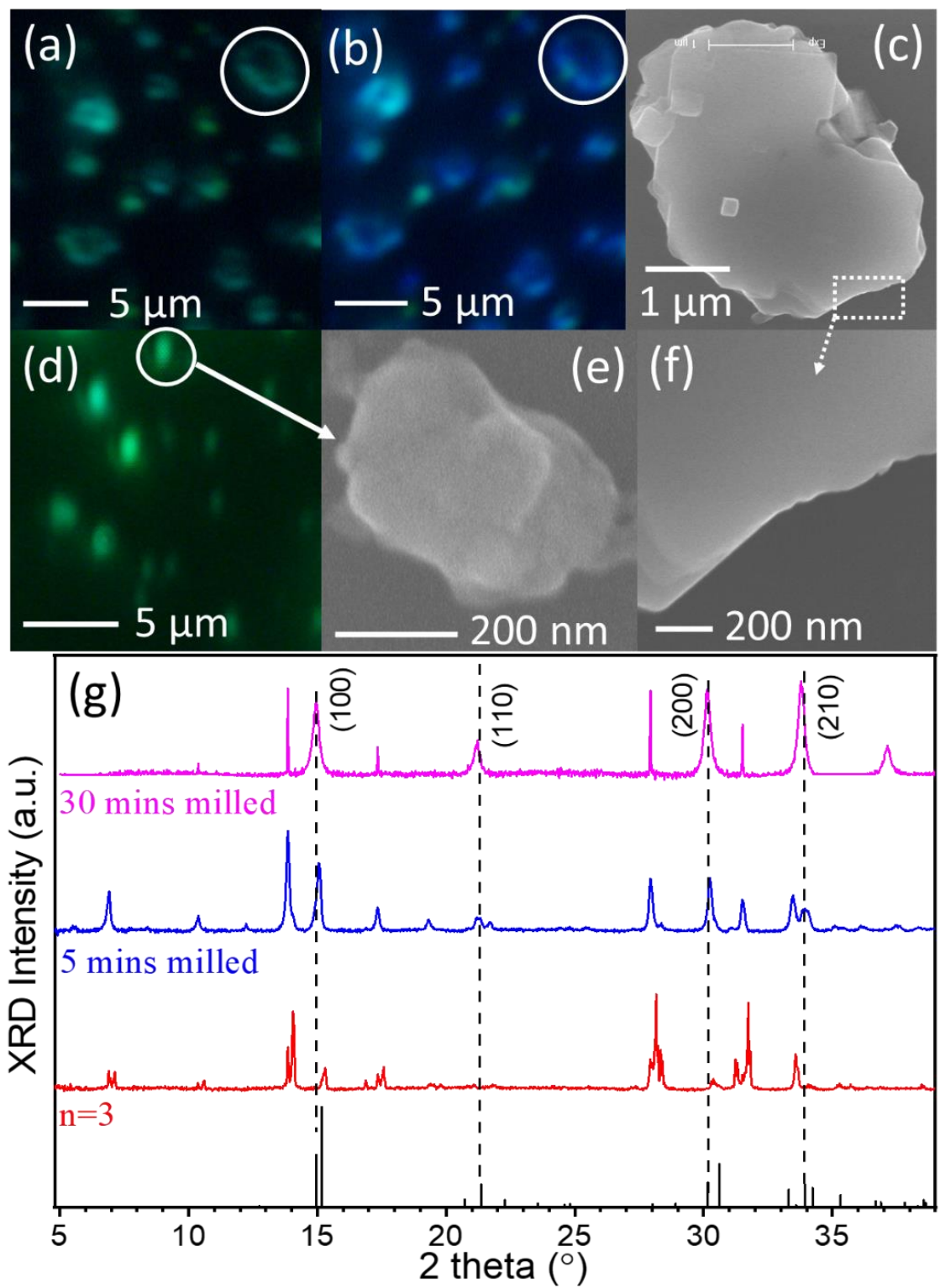

Figure 2. PL images, SEM and XRD of freezer milled $(BA)_2(MA)_2Pb_3Br_{10}$.
(a-b) PL images of 5-minutes milled $(BA)_2(MA)_2Pb_3Br_{10}$ with a 500-nm (a)

and 420-nm (b) long-pass filter. (c, f) SEM of a $(BA)_2(MA)_2Pb_3Br_{10}$ platelet marked in (a) and (b). (d, e) PL and SEM images of 30-mins milled $(BA)_2(MA)_2Pb_3Br_{10}$. (g) Evolution of XRD of $(BA)_2(MA)_2Pb_3Br_{10}$ before and after milling for 5 and 30 minutes.

PL images in figures 2a, 2b and 2d reveal that these physically size-reduced RP perovskites still show green edge emission as initial large-sized crystals. As the milling time increases from 5 to 30 minutes, the average size of perovskite is further reduced to the optical diffraction limit, edge emission becomes difficult to resolve, and PL is dominated by stronger green emission. SEM images in Figure 2c, 2e and 2f show that the surfaces and edges of these size-reduced perovskites are still exceedingly smooth as before. XRD in Fig. 2g confirms that this relative increase in green edge emission is accompanied by the emergence of XRD pattern of 3D $MAPbBr_3$. After milling for 30 minutes, the XRD is dominated by 3D $MAPbBr_3$. Based on the XRD line width, the size of 3D $MAPbBr_3$ is estimated to be ~28 nm. From the strong correlation between green edge emission and XRD of $MAPbBr_3$ we can conclude that the green edge emission in $(BA)_2(MA)_2Pb_3Br_{10}$ is due to $MAPbBr_3$ on the edges. We have also excluded the other two models, i.e., the strain relaxation[15] and chemical difference between Pb and halide[18], because according to these models, $MAPbBr_3$ should not appear in size-reduced RP crystals.

To directly confirm the edge location of $MAPbBr_3$ in the 2D $(BA)_2(MA)_2Pb_3Br_{10}$ platelets and to correlate the property (PL) with structure ($MAPbBr_3$) from the same spot, we turned to Optical Photo-thermal Infrared Spectroscopy (O-PTIR): a non-invasive submicron spatial resolution infrared spectroscopy. This is because the big difference between 2D $(BA)_2(MA)_2Pb_3Br_{10}$ and 3D $MAPbBr_3$ is the lack of BA in $MAPbBr_3$, while IR spectroscopy is very sensitive to the organic molecules. In fact, the absence of BA, and even the change in the ratio of BA to MA has been already detected by FTIR[4, 26, 32]. Fig. 3a shows the FTIR of our own samples: $(BA)_2(MA)_{n-1}Pb_nI_{3n+1}$ for n=1,2,3 and $MAPbBr_3$ for n=∞. It can be seen that with increasing n, the relative intensity of peak located at ~1580 cm$^{-1}$ decreases, and the relative intensity ratio of ~1480 cm$^{-1}$ to ~1580 cm$^{-1}$ increases[4, 32]. Figures 3b-c further show that the intensity change of the ~1580 cm$^{-1}$ peak is accompanied by peak shifting towards 1585 cm$^{-1}$. These changes can be explained by our first-principles calculations (details are given in SI). The vibration that corresponds to 1478 cm$^{-1}$ (CH$_3$ vibration) is strong only in MA$^+$, so intensity of ~1480 cm$^{-1}$ has no change. (BA)$_2$(MA)$_2$Pb$_3$Br$_{10}$ has two IR absorption bands around 1580 cm$^{-1}$ involving NH$_3$ vibrations: one is at 1575 cm$^{-1}$ (BA$^+$) and the other one at 1585 cm$^{-1}$ (MA$^+$). Upon depletion of BA$^+$, the band intensity at 1575 cm$^{-1}$ is decreasing, leading to the reduction of the peak intensity at ~1580 cm$^{-1}$.

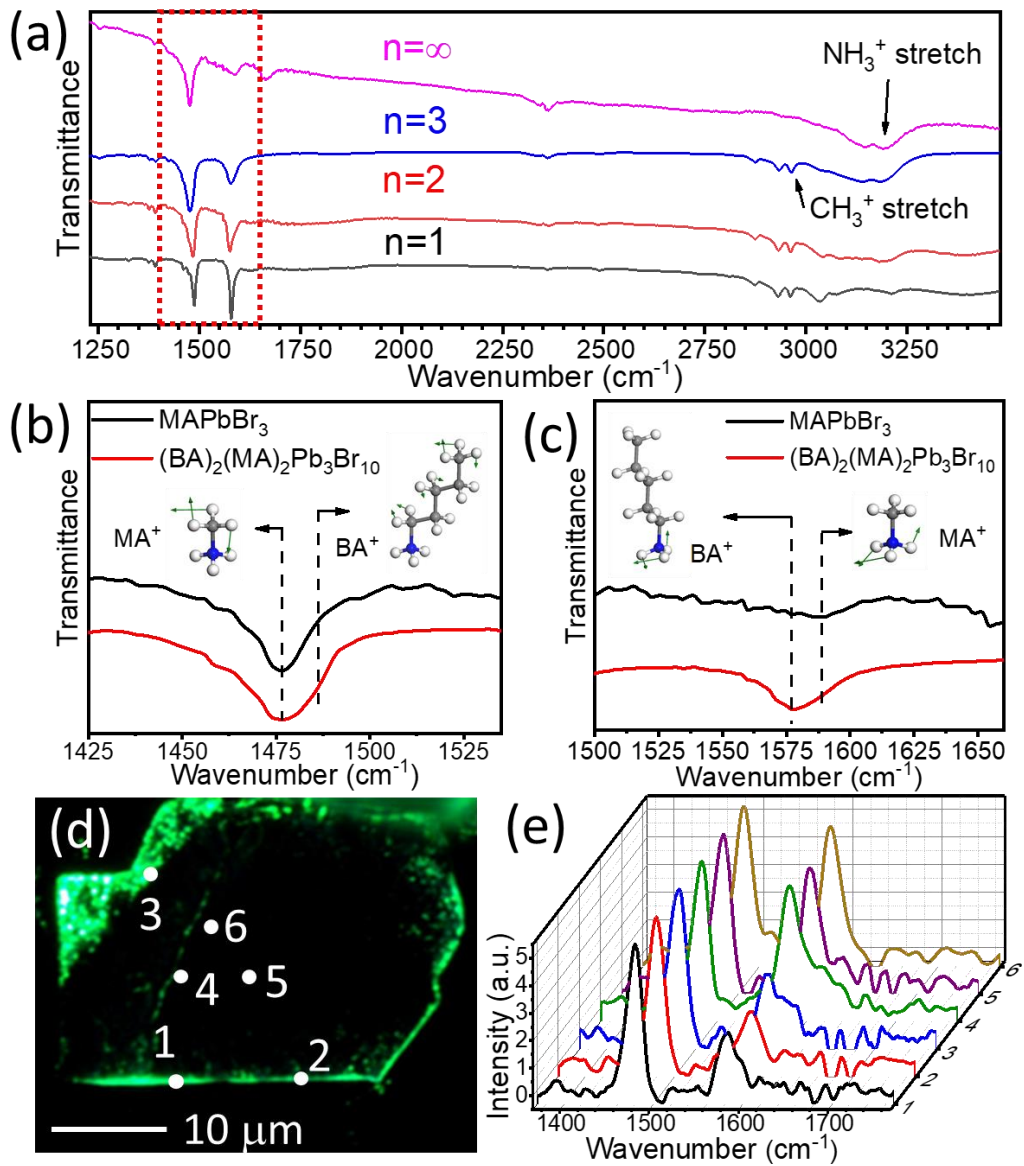

Figure 3. Identification of MAPbBr$_3$ on the edge by IR spectroscopy. (a) FTIR of (BA)$_2$(MA)$_{n-1}$Pb$_n$Br$_{3n+1}$ (n=1, 2, 3, ∞) perovskites. (b-c) FTIR spectra of (BA)$_2$(MA)$_2$Pb$_3$Br$_{10}$ and MAPbBr$_3$ along with the assignment of the bands from vibration calculations of $MA^+$ and $BA^+$ molecules around 1480 cm$^{-1}$ (b) and 1580 cm$^{-1}$ (c); (d) PL images of $(BA)_2(MA)_2Pb_3Br_{10}$. (e) O-PTIR of the center regions and edges indicated in (d).

With the knowledge of spectral difference between $MAPbBr_3$ and $(BA)_2(MA)_2Pb_3Br_{10}$, we used O-PTIR to compare the IR spectra between the edge and the center regions. Figures 3d-e shows three representative spots on the edge and three spots in the center. The peak intensity ratio of ~1480 cm$^{-1}$ to ~1580 cm$^{-1}$ at the edges is larger than that in central regions. This observation is in favor of appearance of BA free structures, of which the only stable one is $MAPbBr_3$.

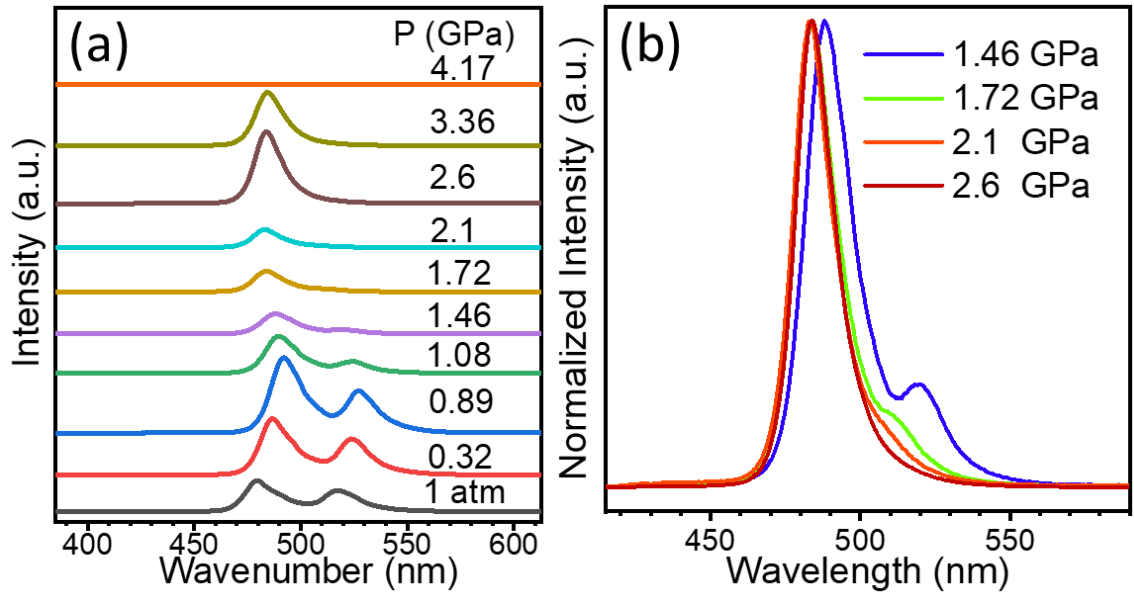

Figure 4. (a) Evolution of PL spectra and peak position of exfoliated $(BA)_2(MA)_2Pb_3Br_{10}$ under hydrostatic pressure. (b) Normalized PL spectra from 1.46 to 2.6 GPa.

The assignment of edge PL to $MAPbBr_3$ can be further confirmed by its spectral response to external hydrostatic pressure[21, 22]. Figure 4 shows the evolution of two PL peaks under increasing pressure. It can be seen that their responses are different, and the response of the edge PL agrees well with that of reported $MAPbBr_3$: its intensity quickly decreases above 1 GPa and diminishes after 2.1 GPa. As a contrast, the PL of the center region remains very strong even under 3 GPa, and only disappears above 4 GPa[33, 34].

The identification of 3D perovskite as photoluminescence center for edge emission in 2D RP perovskite can explain nearly all the reported observations. The explanation of on/off switch of edge PL by BAI and MAI is the most straightforward as MAI can help induce $MAPbBr_3$ on the edge and turn the edge emission on[17]. When exposed to humid air, both BA and MA can interact with $H_2O$ through H bond [32], but as a larger-sized spacer molecule, BA has a much weaker interaction with lead halide octahedron frame, so BA can be more easily coupled with water molecules and gets washed away, resulting in the loss of BA. The availability of MA and loss of BA are necessary for the formation of $MAPbX_3$, where X is halide. That is why the edge emission can be observed when n=2, but not with n=1 unless MA is supplied through MAI. The blue-shift of edge emission compared to the PL peak of bulk 3D perovskites is due to nanometer size of self-formed 3D perovskite nanostructures[21, 22].

The formation of 3D perovskite on the edge of 2D RP perovskite leads to the creation of a lateral 2D/3D perovskite heterostructure. Many other favorable and intriguing properties can be understood through this heterostructure. Because of lower bandgap of 3D compared to 2D RP perovskite, photo-excited excitons can be easily trapped to the edge. The dissociation of the trapped excitons and long carrier lifetime are just unique properties of corresponding 3D perovskite[27, 33, 35, 36]. These interconnected 3D perovskite edges without insulating BA layers are certainly more conductive than the center regions[19].

**Conclusion**

In summary, we employed a series of well-designed experiments and proved that the self-formed 3D perovskite on the edges of 2D RP perovskites is the source for the low energy photoluminescence. This 3D perovskite and the associated 2D/3D perovskite heterostructure are responsible for many beneficial properties that can significantly improve the efficiency of 2D perovskite solar cells. The 3D perovskite is formed after the natural loss or controlled replacement of spacer organic cations with MA. Although lead bromide perovskite is used as an example, this mechanism is applicable to other RP perovskite. 2D/3D vertical[37, 38, 39, 40] and lateral[41] perovskite heterostructures have been used to fabricate high performance and stable solar cells. The identification of edge emission microstructure and understanding of its intriguing properties of 2D/3D heterostructures will be of technological importance for the development of new optoelectronic devices based on the 2D RP hybrid perovskites.